\documentstyle[12pt]{article}

\begin{document}

\begin{titlepage}
\begin{flushright}
OUTP-97-12P\\
hep-th/9703058
\end{flushright}

\begin{center}
{\Large\bf A comment on non-Abelian duality and the strong $CP$ problem}\\
\vspace{1cm}
{\large Jakov Pfaudler\footnote{jakov@thphys.ox.ac.uk}}\\
{\it Department of Physics - Theoretical Physics, University of Oxford\\ 
1 Keble Road, Oxford OX1 3NP, U.K.}\\

\end{center}
\begin{abstract}
It is pointed out that the strong $CP$ problem may have a natural solution
in the context of a recently proposed dualized version of the Standard 
Model where Higgs fields and generations emerge naturally. Although 
fermions have finite pole-masses, the fermionic mass matrix itself is factorizable (having only one nonzero eigenvalue) to all orders in perturbation theory thus allowing one to perform a chiral transformation $\psi\rightarrow\psi'=e^{-i\gamma_{5}\alpha}\psi$ and to rotate the $\theta$-angle to zero.
\end{abstract}

\end{titlepage}

\clearpage

\baselineskip 16pt
One of the unsolved problems within the Standard Model is the strong 
$CP$ problem: Because of nonperturbative effects, the QCD-Lagrangian contains 
a $\theta$-term which violates $P$ and $T$ and thus $CP$. The physically relevant parameter is the effective $\bar{\theta}$-angle in the basis 
where the renormalized quark mass matrix $m_R$ is real and diagonal. 
It is defined in an arbitrary basis by:

\begin{equation}
\bar{\theta}=\theta_{QCD}+\theta_{QFD},
\end{equation}

\noindent where $\theta_{QFD}=\arg\det m_R$.

From measurements of the neutron electric dipole moment however, 
it is known that strong interactions conserve $CP$ rather well. 
These constraints give an upper bound for the $\bar{\theta}$ parameter 
of 

\begin{equation}
\bar{\theta}<10^{-9}. \nonumber
\end{equation}

This is the strong $CP$ problem: if $\bar{\theta}$ is a genuine 
parameter of QCD, why is it so small?
A number of mechanisms have been suggested to resolve the $CP$ 
problem {\cite{review} most of which involve the existence of an 
additional pseudoscalar field, the $axion$, first introduced by 
Peccei and Quinn \cite{PecQ}. Experimental searches and 
astrophysical arguments have shown that the ``axion hypothesis" can 
be ruled out unless the axion only interacts weakly and is very light 
with an upper bound for the mass of $10^{-3}eV$ \cite{axmass}. So far 
there is no experimental evidence for the existence of axions.

It is well known that the strong $CP$ problem could be avoided if at 
least one quark had a zero mass. Transforming to the basis where 
$m_R$ is real and diagonal we then have $\theta_{QFD}$=0 
and $\bar{\theta}=\theta_{QCD}$. A chiral rotation on the massless 
quark

\begin{equation}
 \psi\rightarrow\psi'=e^{-i\gamma_{5}\alpha}\psi
\label{chirrot}
\end{equation}

\noindent would then 
change the fermionic measure in the path integral 
\cite{fuji}:

\begin{equation}
d\mu=\prod_{i}{\cal D}\bar{\psi}_{i}{\cal D}\psi_{i}\rightarrow d\mu'=e^{i\Delta} d\mu,
\end{equation}

\def\Fp1{/\hspace{-0.65em}D}

\noindent where

\begin{equation}
\Delta=\frac{e^2\alpha}{8\pi^2}\int d^{4}x\,  tr F_{\mu\nu}\, ^{*}F^{\mu\nu}.
\end{equation}

It is then possible to absorb the $\theta$ dependence into the 
fermionic sources and obtain $\bar{\theta}=0$ by a suitable choice 
of $\alpha$ since the Green's functions are defined in the sourceless 
limit. We thus see that in the presence of at least one massless quark 
all $\theta$-worlds are physically equivalent and strong interactions 
are invariant under $CP$. The problem with this solution is, of course, 
that none of the quarks seem to have a zero mass

Recently a ``dualized version" of the Standard-Model has been 
proposed \cite{hm1} which is based on a non-Abelian duality introduced 
in \cite{hm2}. Let us briefly recall its main features. As shown in 
\cite {hm2} for a gauge theory with semisimple gauge group $G$ the 
symmetry is enlarged to $G\times\tilde{G}$ where $\tilde{G}$, the dual 
of $G$ has opposite parity to $G$. A dual potential $\tilde{A}_\mu$ 
which transforms under $\tilde{G}$ is constructed which couples to 
the monopoles of the theory. As elaborated in \cite{hm1}, $\tilde{G}$ 
does not represent an additional degree of freedom: following 
't Hooft's arguments \cite{tho} $\tilde{G}$ has to be confined when $G$ 
is broken and vice versa. In the context of the Standard Model, colour 
is confined and weak isospin is broken. Although the non-Abelian 
duality transformation is expressed in terms of loop-variables and 
reduces to the Hodge-* transformation only in the Abelian case, it 
yields nevertheless an explicit transformation relating $F^{\mu\nu}$ 
to $\tilde{F}^{\mu\nu}$. In this transformation a central role is 
played by the map $\omega(x):\, G\rightarrow\tilde{G}$ relating 
the (conjugate) fundamental representation of $G$ to the 
corresponding fundametal representation of $\tilde{G}$ at each point 
in space-time. This map $\omega$  is then promoted to a triplet of 
Higgs fields $\phi^{(a)}$ whose vacuum expectation values constitute 
an orthogonal frame in internal symmetry space. These vacuum 
expectation values break the colour-$\widetilde{SU(3)}$ symmetry of 
the Lagrangian. This is in accordance with 't Hooft's argument: 
since colour-$SU(3)$ is confined, colour-$\widetilde{SU(3)}$ has to 
be spontaneously broken. The three generations of fermions are 
then interpreted as spontaneously broken dual colour. Hence quarks 
carry both a colour index and a generation index (dual colour) and are 
thus dyons\footnote{Only the left handed fermions carry the dual 
colour charge defined in the topological manner of \cite{hm4}, the 
right handed fermions are dual colour neutral but there are three of 
them carrying a label $[b]$, $ b=1,\, 2,\, 3$}. The Higgs fields carry 
dual colour but no colour.

Defining $\tilde{\phi}=\sum_{(a)} \tilde{\phi}^{(a)}$ and a row vector 
of Yukawa couplings $Y=\left(a,b,c\right)$ where $a$, $b$, $c$ are 
complex parameters, the Yukawa coupling of fermions and Higgs fields 
is written as

\begin{equation}
{\cal L}_{Y}=(\bar{\psi}_L)^{+}\tilde{\phi}\; Y\; (\psi_R) + \mbox{h.c.}.
\end{equation}

As usual, the mass matrix for the fermions is then obtained by inserting 
for $\tilde{\phi}$ its vacuum expectation value which is taken to 
be

\begin{equation}
 \tilde{\phi}_V = \left( \begin{array}{c}
   x\\y\\z \end{array} \right).
\label{phitvacu}
\end{equation}

\noindent Note that the vacuum expectation value of $\phi$ is the same 
for $u$- and $d$-type quarks but the Yukawa coupling $(a,b,c)$ is 
different for different types. This leads to a factorizable mass matrix 
of rank one: 

\begin{equation}
m=\left( \begin{array}{ccc} ax & ay & az \\
                           bx & by & bz \\
                           cx & cy & cz \end{array}\right) = \left( \begin{array}{c}
   x\\y\\z \end{array} \right)\left(a,b,c\right).
\end{equation}

\noindent It can be diagonalized to:

\begin{equation}
m_D=\left( \begin{array}{ccc} 0 & 0 & 0 \\
                           0 & 0 & 0 \\
                           0 & 0 & \rho\zeta \end{array}\right),
\end{equation}
\noindent where $\zeta=\sqrt{x^2+y^2+z^2}$ and $\rho=\sqrt{|a|^2+|b|^2+|c|^2}$. 
The unitary matrix $U$ that diagonalizes $m$ depends only on the 
vacuum expectation value of the Higgs fields $(x,y,z)$ and is thus the 
same for $u$-and $d$-type quarks. Hence the CKM matrix is unity at 
tree level and only the heaviest member in each family has a nonzero 
mass. This is a unique feature of the dualized Standard Model 
contrasting with the usual Higgs mechanism where the Yukawa coupling 
is given by a complex (3$\times$3)-matrix which has a priori no 
zero eigenvalue and the CKM matrix is an empirical quantity depending
on four parameters to be determined experimentally.

The crucial point is, that the (renormalized) mass matrix $m_R$ remains 
factorizable, having thus only one nonzero eigenvalue, to all orders 
in perturbation theory. Hence to all oders in perturbation theory we 
are able to perform the chiral rotation (\ref{chirrot}) necessary 
for rotating the $\theta$-angle to zero. As shown in \cite{hm1} loop
corrections of the kind described by Weinberg \cite{wei} will rotate 
the left hand factor in the mass matrix:

\begin{equation}
m \rightarrow m_R = \left( \begin{array}{c}
   x_1 \\ y_1 \\ z_1 \end{array} \right) (a, b, c),
\end{equation}

\noindent where in general the left hand factor $(x_1, y_1, z_1)$ will 
be different for {\it u}-and {\it d}-type quarks. This induces 
a complex CKM matrix different from the identity.

This factorizable form of $m_R$, however does not imply that the two 
lower generations have vanishing pole-masses. Hence it does not contradict 
the empirical fact of non zero pole-masses for all generations. The 
loop corrections, apart from rotating the left hand factor above, induce a 
scale dependence of the mass matrix and let the eigenvalues run via 
the renormalization group equation. However, since the mass matrix 
remains factorizable and of rank one at every scale, it is not 
immediately clear how to 
define the mass of the lower generations. Chan and Tsou suggest 
the following procedure. Recall first that the usual definition (in 
the $\bar{MS}$-scheme) of the mass of a particle is \cite{databook}:

\begin{equation}
m_{\bar{MS}}(m_Q)=m_Q. \label{runmass}
\end{equation}

At every scale diagonalize the mass matrix and obtain one 
nonzero eigenvalue. The scale at which this eigenvalue equals the 
scale itself as in (\ref{runmass}) is designated as the physical mass of 
the heaviest particle. The corresponding eigenvector is identified with 
the state vector of this particle. The other two eigenvalues are zero 
at this scale, but this does not correspond to zero masses for the 
lower generations since these masses have to be evaluated at a 
different scale, according to (\ref{runmass}). If one now considers 
the submatrix $m^{(2)}_R$ of $m_R$ for the two lower generations one 
can again diagonalize and evaluate it at every scale. As explained 
in \cite{hm1}, this ``running down" will de-diagonalize the submatrix 
$m^{(2)}_R$. Here it is important to note that also the mass of the heaviest 
generation member will change when running the mass matrix. 
Crudely speaking, the highest generation particle ``leaks" a fraction 
of its mass to the other generations. The mass of the second generation 
is then assigned to be the eigenvalue of $m^{(2)}_R$ at 
which (\ref{runmass}) holds. The corresponding eigenvector is the 
physical state-vector of the second generation particle. The mass of 
the lowest generation is then defined in the same manner. 

We thus see that this scheme allows zero masses in the 
Lagrangian without spoiling the non-vanishing of all quark 
(pole-)masses. The crucial fact is the factorizability of the mass matrix 
to all orders in perturbation theory. Hence it is possible to assign 
nonzero physical masses to all quarks while simultaneously allowing $\theta_{QCD}$ to be rotated away thereby restoring $CP$-invariance 
in the strong interaction and curing the strong $CP$ problem\footnote{Note however, that the observed $CP$-violation of the weak interaction is 
accomodated for by a complex phase in the CKM matrix after loop corrections.}. \\\\ 
\newpage
\noindent{\Large\bf Acknowledgements}\\

I am thankful to Chan Hong-Mo, Tsou Sheung Tsun and to Jose Bordes for many 
helpful discussions and encouragement. This work was supported 
by the German Academic Exchange Service (DAAD).

\end{document}